\documentclass[aps,prb,twocolumn,floatfix]{revtex4}
\usepackage{amsmath,amssymb,epsfig,psfig}

\begin{document}


\title{Combined local-density and dynamical mean field theory
calculations \\ for the compressed lanthanides Ce, Pr, and Nd}

\author{A. K. McMahan}

\affiliation{Lawrence Livermore National Laboratory, University
of California, Livermore, CA 94550}

\begin{abstract}

This paper reports calculations for compressed Ce ($4f^1$), Pr
($4f^2$), and Nd ($4f^3$) using a combination of the local-density
approximation (LDA) and dynamical mean field theory (DMFT),
or LDA+DMFT.  The $4f$ moment, spectra, and the total energy
among other properties are examined as functions of volume and
atomic number for an assumed face-centered cubic (fcc) structure.
These materials are seen to be strongly localized at ambient
pressure and for compressions up through the experimentally
observed fcc phases ($\gamma$ phase for Ce), in the sense of
having fully formed Hund's rules moments and little $4f$ spectral
weight at the Fermi level.  Subsequent compression for all three
lanthanides brings about significant deviation of the moments from
their Hund's rules values, a growing Kondo resonance at the Fermi
level, an associated softening in the total energy, and quenching
of the spin orbit since the Kondo resonance is of mixed spin-orbit
character while the lower Hubbard band is predominantly $j\!=\!5/2$.
While the most dramatic changes for Ce occur within the two-phase
region of the $\gamma$--$\alpha$ volume collapse transition, as
found in earlier work, those for Pr and Nd occur within the volume
range of the experimentally observed distorted fcc (dfcc) phase,
which is therefore seen here as transitional and not part of the
localized trivalent lanthanide sequence.  The experimentally observed
collapse to the $\alpha$-U structure in Pr occurs only on further
compression, and no such collapse is found in Nd.  These lanthanides
start closer to the localized limit for increasing atomic number,
and so the theoretical signatures noted above are also offset to
smaller volume as well, which is possibly related to the measured
systematics of the size of the volume collapse being 15\%, 9\%,
and none for Ce, Pr, and Nd, respectively. \\

\end{abstract}



\maketitle

\section{Introduction}

A number of the trivalent lanthanides undergo first order phase
transitions under pressure that are characterized by unusually large
volume changes, Ce (15\%),\cite{KoskimakiCe78,OlsenCe85,VohraCe99}
Pr (9\%),\cite{MaoPr81,ZhaoPr95,ChesnutPr00,BaerPr03}
 Gd (5\%),\cite{HuaGd98} and Dy (6\%).\cite{PattersonDy04}
These ``volume collapse'' transitions are believed
to be caused by changes in the degree of $4f$ electron
correlation from strongly correlated (localized) at pressures
below the transitions to more weakly correlated (itinerant)
above.\cite{Benedict93,Holzapfel95,JCAMD} Associated with these two
regimes are differences in physical properties, with high-symmetry
crystal structures below the collapse transitions characteristic of
metals without $f$ electrons, and low-symmetry early-actinide-like
structures at pressures above, suggesting a greater participation
of $f$ electrons in the bonding.  The magnetic susceptibility is
of Curie-Weiss character in the localized regime characteristic of
Hund's rules $4f$ moments, and is believed to become temperature
independent similar to Pauli paramagnetism in the itinerant regime.
The latter behavior has actually been observed only for the collapsed
$\alpha$ phase of Ce, but is inferred for the other lanthanides
based on analogies to the actinides.

In contrast to these apparently general trends, it has been recently
shown that Nd reaches the characteristic itinerant $\alpha$-U
structure in a relatively continuous fashion without any phase
transitions of unusually large volume change,\cite{ChesnutNd00}
while in an analogous actinide system, Am is noted to undergo two
collapse transitions.\cite{HeathmanAm00} It would therefore appear
that the evolution from localized to itinerant character in these $f$
electron metals occurs more generally in a continuous manner which
may or may not be accelerated by one or more large-volume-change
collapse transitions.

It was demonstrated many years ago that the volume collapse
transitions in the lanthanides and actinides do not involve
the promotion of electrons from $f$ to other states, and it
was argued that these metals were instead undergoing Mott
transitions in their respective $4f$ and $5f$ electron
systems.\cite{Johansson74,Johansson75} Local density
approximate (LDA) calculations modified to enable orbital
polarization,\cite{Eriksson90,SoderlindPr02,SvanePr97} to include the
self-interaction correction,\cite{SvanePr97,SvaneCe94,SzotekCe94}
and to explicitly incorporate the Hubbard repulsion
$U_f$,\cite{Sandalov95,Shick01} appear to provide such transitions
for Ce and Pr.  They describe large-volume magnetically-ordered
solutions with Hubbard split $4f$ bands which become unstable under
compression to paramagnetic solutions where these bands are grouped
near the Fermi level.  It has been suggested, however, that the
rather abrupt delocalization of these modified-LDA transitions is
an artifact of the static-mean-field nature of these techniques,
and that a correlated solution would allow a continuous transfer of
$f$ spectral weight between the Hubbard side bands and the Fermi
level over an extended volume range.\cite{Held00} This is the
case for the Kondo-volume collapse model of the Ce transition,
based on a many-body solution of the Anderson impurity model,
which identifies Kondo-like screening of the local moment
as the driving force for the transition.\cite{KVC,KVC2,KVC3}
More realistic calculations which combine LDA with correlated
Dynamical Mean Field Theory (DMFT),\cite{DMFT1,DMFT2} so called
LDA+DMFT,\cite{LDADMFT1,LDA++,LDADMFT3}  generally corroborate these
characteristics for Ce,\cite{Zoelfl01,Held01,McMahan03,Haule05}
and the same technique has been applied to the
actinides.\cite{Savrasov01}

Beyond the specific case of Ce, a more general understanding of
the lanthanide volume-collapse transitions or their absence may
come from putting them in context of the extended evolution from
localized to itinerant character which occurs as these materials
are compressed.  As a step in this direction, the present paper
reports LDA+DMFT calculations over a wide range of volume for Ce
($4f^1$), Pr ($4f^2$), and Nd ($4f^3$), the last as mentioned
being notable for the absence of a volume collapse transition.
These calculations have been carried out as in earlier work on
Ce,\cite{Held01,McMahan03} except that the spin-orbit interaction
is now incorporated.  They also assume a face centered cubic (fcc)
structure throughout, although LDA estimates of the structural
energy differences suggest these are significantly smaller than the
relevant contributions to the correlation energy, which may then
be discussed in terms of their volume dependence as more important
leading order effects.

The trivalent lanthanides follow a general structural sequence
under pressure, namely hcp $\rightarrow$ Sm-type $\rightarrow$ dhcp
$\rightarrow$ fcc $\rightarrow$ dfcc, where these abbreviations are
for hexagonal close packed (hcp), double hcp (dhcp), and distorted
fcc (dfcc).\cite{Benedict93,Holzapfel95,JCAMD} These are all just
different stacking variants within the family of close packed
structures, except for dfcc, a soft L-point phonon distortion
of fcc,\cite{HamayaPr93} a phase which only Ce does not assume.
The heaviest lanthanides traverse the full sequence, whereas the
lighter ones begin part way, e.g., Ce, Pr, and Nd are all dhcp
at ambient conditions, although fcc Ce is also metastable there.
Since the same sequence is seen in Y which has no nearby $f$
states,\cite{Vohra81} and is theoretically understood to depend
on the $5d$ occupancy,\cite{Duthie77} it would appear to have no
relation whatsoever with the $f$ electrons, which are then presumed
to be fully localized.  Indeed, in the present work we find evidence
for fully formed Hund's rules moments and little $4f$ spectral weight
at the Fermi level throughout the experimentally observed stability
field of the fcc phases ($\gamma$ for Ce), and for lower pressures.

Further compression takes Ce through a 15\% volume collapse from the
$\gamma$ (fcc) to the $\alpha$ (also fcc) phase, while Pr undergoes
a 9\% collapse from the dfcc to an $\alpha$-U structure, and Nd
passes from the dfcc through two other low-symmetry structures
before also reaching $\alpha$-U, however, without any large
volume changes.  The present work finds a variety of signatures
to accompany this compression regime: substantial deviation of the
moments away from their Hund's rules values, rapid growth in $4f$
spectral weight at the Fermi level (the Kondo resonance) at the
expense of the Hubbard side bands, an associated softness in the
total energy, and quenching of the spin orbit due to the mixed-$j$
character of the Kondo resonance versus the predominant $j\!=\!5/2$
lower Hubbard band.  The most dramatic changes in these signatures
for the case of Ce {\it coincide} with the two-phase region in the
$\gamma$--$\alpha$ transition.  This behavior was observed in earlier
work on Ce,\cite{Zoelfl01,Held01,McMahan03} and is consistent with
the Kondo volume collapse scenario.\cite{KVC,KVC2,KVC3}

In regard to signatures in the total energy, a unique feature
of the Kondo volume collapse model is the assertion that
the $\gamma$--$\alpha$ transition is driven by a rapid drop
in entropy from the $\gamma$ to the $\alpha$ side, reflecting
screening of the $4f$ moments.  The transition therefore need not
occur at $T\!=\!0$,\cite{KVC,KVC3} as has been documented for Ce
alloys,\cite{Thompson83} implying a featureless $T\!=\!0$ total
energy.  Nevertheless, due to the interrelationships between the
different thermodynamic functions, and even though it is the free
energy which actually determines the phase transition, the {\it
finite-}$T$ total energy should show an associated softening
in the vicinity of the phase transition, and is therefore a useful
diagnostic.\cite{softenergy} 

The present work finds the same signatures for compressed Pr and
Nd, however, their most dramatic change takes place over the volume
range where the dfcc phase is experimentally observed, suggesting
this phase is of transitional character and not the end member of
the localized trivalent lanthanide sequence as has been assumed.
Only on further compression is Pr observed to collapse into the
$\alpha$-U structure, and as noted Nd has no collapse.  An important
difference among the lanthanides is that they become more localized
with increasing atomic number, due to the increased but incompletely
screened nuclear charge.  This is evident in the present work by the
offset of the concurrent correlation-related signatures to smaller
volumes from Ce to Pr and then to Nd.  One consequence is that the
softening effect in the energy associated with the growing Kondo
resonance must compete with the remaining, dominant part of the
energy which has a curvature that grows ever larger with decreasing
volume.  While a region of negative bulk modulus is required for an
isostructural transition as in Ce, it may be more generally that a
low bulk modulus favors larger volume changes in structural phase
transitions, which would then be consistent with the decreasing
size of the collapse from 15\% (Ce), to 9\% (Pr), to none (Nd).
The observed collapse transitions in Gd and Dy would then appear
inconsistent; however, these involve filling of the $j\!=\!7/2$
subshell which complicates matters.

In the remainder of this paper, the theoretical methods are reviewed
in Sec.~\ref{theorysec}, numerical results for compressed Ce, Pr,
and Nd are presented in Sec.~\ref{resultsec}, and then a summary
and discussion is given in Sec.~\ref{summarysec}.

\section{Theoretical methods}
\label{theorysec}

The results in this paper unless otherwise indicated have been
obtained by the LDA+DMFT(QMC) method, which refers to the merger
of the local density approximation (LDA) with dynamical mean field
theory (DMFT),\cite{DMFT1,DMFT2} in order to create a composite
technique\cite{LDADMFT1,LDA++,LDADMFT3} which rigorously treats
onsite electron correlations and yet retains the material realism and
specificity of the LDA.  An essential component of the DMFT method
is solution of an auxiliary impurity problem which is achieved here
by a quantum Monte Carlo (QMC) algorithm.\cite{Hirsch86,QMC1bnd}
The present calculations have been carried out as described in
previous work for Ce,\cite{Held01,McMahan03} except that the
spin-orbit interaction has now been added and a broader set of
materials considered.  In the remainder of this section we give
a brief review of the method, various computational details, and
calculation of the moment.

\subsection{The LDA+DMFT method}
\label{ldadmftsec}

The LDA contribution to the LDA+DMFT method is to provide an
effective Hamitonian which includes all valence electron degrees
of freedom,
\begin{eqnarray}
H = &&\sum_{{\bf k},ljm,l^\prime j^\prime m^\prime} 
[H^0_{\rm LDA}({\bf k})]_{ljm,l^\prime j^\prime m^\prime}
\,\hat{c}^\dagger_{{\bf k}\,ljm}
\hat{c}^{ }_{{\bf k}\,l^\prime j^\prime m^\prime}
\nonumber \\
&&+\; \frac12 \, U_f \!\!\! \sum_{{\bf i},j m,j^\prime m^\prime}
\!\!\!\!\!\!\! ^{'} \,
\hat{n}_{{\bf i}fjm}\, \hat{n}_{{\bf i}fj^\prime m^\prime}.
\label{hameqn}
\end{eqnarray}
Here, {\bf k} are Brillouin zone vectors, ${\bf i}$ are lattice
sites, $l$ is the orbital angular momentum, $j$ is the total
angular momentum ($l\!\pm\!\frac12$ except just $\frac12$ for
$l\!=\!0$), $m=-j,-j\!+\!1,\cdots, j$, $\hat{n}_{{\bf i}fjm} \equiv
\hat{c}^\dagger_{{\bf i}fjm} \hat{c}^{ }_{{\bf i}fjm}$, and the
prime signifies $jm \neq j^\prime m^\prime$.  The relevant valence
states for the present lanthanide case are $6s$, $6p$, $5d$, and
$4f$, and so the matrices $H^0_{\rm LDA}({\bf k})$ are $32\times 32$.

As described elsewhere,\cite{JCAMD} the $H^0_{\rm LDA}({\bf
k})$ are orthogonalized one-electron Hamiltonian matrices
obtained from converged linear muffin-tin orbital LDA
calculations,\cite{LMTO1,LMTO2} in which the $4f$ site
energies are shifted so as to avoid double counting the $f$-$f$
Coulomb interaction $U_f$ which is explicitly incorporated into
Eq.~(\ref{hameqn}).  The spin-orbit interaction was included as a
perturbation,\cite{LMTO1} with the spin-orbit coupling parameters
$\xi_l$ kept fixed across the bands, although evaluated at the most
important energies, namely at the respective centers of gravity of
the occupied state density for each orbital type.  The resultant
$H^0_{\rm LDA}({\bf k})$ matrices were calculated over a grid of
volumes for Ce, Pr, and Nd in an assumed face centered cubic (fcc)
structure, and companion LDA constrained-occupation calculations
used to get the volume and material dependent Coulomb interactions
$U_f$.\cite{JCAMD}

Dynamical mean field theory assumes a local or {\bf k}-independent
self-energy $\Sigma(i\omega)$, which together with the one-body
part of Eq.~(\ref{hameqn}) provides the lattice Green function,
\begin{equation}
G_{\bf k}(i\omega)=\left[ \;i\omega +\mu
-H_{{\rm LDA}}^{0}({\bf k})
-\Sigma(i\omega)\right]^{-1} .
\label{latgrneqn}
\end{equation}
This is solved in tandem with an auxiliary impurity problem defined
by a similar Dyson-like equation with the same local self energy.
\begin{equation}
G^{\rm imp}(i\omega)= [{\cal G}(i\omega)^{-1}-\Sigma(i\omega)]^{-1} \, .
\label{impgrneqn}
\end{equation}
The idea is to guess an initial $\Sigma=\Sigma^{(1)}$, calculate
the lattice Green function $G_{\bf k}$ from Eq.~(\ref{latgrneqn}),
identify its {\bf k} average with $G^{\rm imp}$, and then find the
noninteracting or bath Green function ${\cal G}={\cal G}^{(1)}$
for the impurity problem from Eq.~(\ref{impgrneqn}).  The impurity
problem is completely defined by ${\cal G}$, which describes
its one-body part, together with the Coulomb interaction for its
single site as in Eq.~(\ref{hameqn}) without the {\bf i} index.
It may be solved exactly to within statistical uncertainties by
QMC techniques.\cite{Hirsch86} This gives a new $G^{\rm imp}$ which
together with the input ${\cal G}^{(1)}$ defines a new self-energy
$\Sigma=\Sigma^{(2)}$ again via Eq.~(\ref{impgrneqn}).  The cycle
is then repeated by starting anew with $\Sigma=\Sigma^{(2)}$ in
the lattice Green function Eq.~(\ref{latgrneqn}), and so on until
self consistency.

Quantities of interest such as the number $n_j$ of $j\!=\!5/2$ or
$7/2$ $4f$ electrons may be obtained from the lattice Green function
\begin{equation}
n_j \! =\! \frac{T}{N} \sum_{n {\bf k}}
{\rm Tr}_{fj} \left[G_{\bf k}(i\omega_n)\right] e^{i\omega_n 0^+} \, ,
\label{njeqn}
\end{equation}
where the sums are over {\bf k} vector and Matsubara frequency
$\omega_n=(2n\!+\!1)\pi T$, and the trace is over those $2j\!+\!1$
$4f$ states of type $j$. The total number of $4f$ electrons is
then $n_f=n_{5/2} +n_{7/2}$.  Similarly the energy per site for
the effective Hamiltonian in Eq.~(\ref{hameqn}) is
\begin{equation}
E_{\rm DMFT}\! =\! \frac{T}{N} \sum_{n {\bf k}}
{\rm Tr}\left[{ H}^0_{\rm LDA}({\bf k}) { G}_{\bf k}(i\omega_n)\right]
e^{i\omega_n 0^+} + U_f \, d \, ,
\label{edmfteqn}
\end{equation}
which involves the double occupation
\begin{equation}
d \!=\! \frac{1}{2}
\sum_{j m,j^\prime m'}^{\;\;\;\;\;\;\;\;\prime} \langle
\hat{n}_{fjm}\, \hat{n}_{fj^\prime m^\prime}\rangle \,  ,
\label{dbleqn}
\end{equation}
as discussed elsewhere.\cite{Held01,McMahan03} In principle this is
an average over the sites {\bf i}, although in practice we obtain
this quantity form the QMC impurity problem.

To evaluate the total LDA+DMFT energy $E_{\rm tot}$ including
all core and outer electrons, we add a correction to the
paramagnetic all-electron LDA energy $E_{\rm LDA}$
\begin{equation}
E_{\rm tot}=E_{\rm LDA}+E_{\rm DMFT}-E_{\rm mLDA} \, ,
\label{etoteqn}
\end{equation}
which consists of the DMFT energy $E_{\rm DMFT}$ from
Eq.~(\ref{edmfteqn}) less an LDA-like solution of the model
Hamiltonian Eq.~(\ref{hameqn}), thus ``model LDA'' or $E_{\rm mLDA}$.
To be more accurate, our LDA total energy $E_{\rm LDA}$ is scalar
relativisitc and so omits spin orbit, and therefore the correction
$E_{\rm mLDA}$ should be and is obtained from a modification
of Eq.~(\ref{hameqn}) with spin orbit removed.  Using this
modified effective Hamiltonian, $E_{\rm mLDA}$ is determined by a
self-consistent solution of Eqs.~(\ref{latgrneqn}) and (\ref{njeqn})
for $n_f=n_{5/2}+n_{7/2}$ taking a self-energy $\Sigma_{\rm mLDA}
= U_f(n_f-\frac12)$. From this, the kinetic energy is calculated
by the first term of Eq.~(\ref{edmfteqn}) and the potential energy
by $\frac12 U_f n_f(n_f-1)$.

The simple form of Eq.~(\ref{etoteqn}) is a reminder of the
still fairly new and evolving nature of the LDA+DMFT method.
There are recent discussions of the formal context of the
method,\cite{Savrasov04} as well as some sense that a GW platform
in place of the LDA may allow for more natural integration
with DMFT.\cite{Biermann03,Sun04} Mutual self-consistency
between the LDA and DMFT parts has also been stressed, i.e.,
feeding a different DMFT value of $n_f$ back into a new LDA
calculation.\cite{Savrasov01,Savrasov04} This is impractical in the
present case given the expense of the QMC solutions, although we
do provide a perturbative change in $n_f$ in going from the LDA to
the DMFT, and this may well be sufficient.  A step towards answering
many of these questions is simply to provide more quantitative tests
against experiment of LDA+DMFT implementations with clearly stated
approximations, and this is the spirit of the present work.

\subsection{Computational details}
\label{cmpdtlsec}

While the matrices $G_{\bf k}$, $H^0_{\rm LDA}({\bf k})$, and
$\Sigma$ in Eq.~(\ref{latgrneqn}) are all $32\times 32$, it is
customary to take $\Sigma$ non-zero only within the interacting or
$14\times 14$ $f$-$f$ block, which permits the same reduction in the
impurity problem itself.  Even so, one must still determine seven
distinct functions of $\omega$ to fill out $\Sigma(i\omega)$ even
for the full cubic point group,\cite{cubirr} and more otherwise.
On the other hand, if we ignore crystal field, then the $f$-$f$
self-energy becomes
\begin{equation}
\Sigma_{jm,j^\prime m^\prime}(i\omega) = \Sigma_j(i\omega)
\delta_{j j^\prime} \delta_{m m^\prime} \, ,
\label{slfeeqn}
\end{equation}
and we need only find functions for the two spin-orbit states
$j\!=\!5/2$ and $7/2$.  This is an excellent approximation over most
of the volume range studied here, including the volume-collapse
regions, where the crystal field splitting ranges from about
1 to 10\% of the spin-orbit splitting.  At the very smallest
volumes considered here, this ratio approaches 50\%, however,
it is to be emphasized this omission is only in the self-energy,
and all such effects are reflected in $H^0_{\rm LDA}({\bf k})$
where hybridization dominates anyway.  In practice we input a bath
Green function of similar structure, ${\cal G}_j\delta_{j j^\prime}
\delta_{m m^\prime}$, into the QMC and then improve statistical
uncertainties by averaging over the six $j\!=\!5/2$ and eight $7/2$
states in determining the output $G^{\rm imp}_j$.

It was noted previously that the DMFT(QMC) iterations could be
greatly accelerated by separately converging the leading, constant,
Hartree-Fock-like part of the self-energy outside of the QMC part of
the cycle.\cite{McMahan03} For the present case this constant part is
\begin{equation}
\Sigma^{\rm (0)}_j = [n_f - n_j/(2j\!+\!1)]U_f \, .
\label{sig0eqn}
\end{equation}
Therefore after each QMC cycle we extract just the frequency
dependent part of the self-energy
\begin{equation}
\Delta\Sigma_j(i\omega) = \Sigma^{\rm QMC}_j(i\omega) -
\Sigma^{\rm (0)}_j(\{n^{\rm QMC}_j\}) \, ,
\label{delsigeqn}
\end{equation}
using here the QMC $n_j$ values
\begin{equation}
n^{\rm QMC}_j=(2j\!+\!1)[1\!+\!\tilde{G}^{\rm imp}_j(0^+)] \, ,
\label{nfqmcgeqn}
\end{equation}
where $\tilde{G}^{\rm imp}_j(\tau)$ is the imaginary
time Fourier transform of $G^{\rm imp}_j(i\omega)$.
Then with $\Delta\Sigma_j(i\omega)$ held fixed, we
construct a new self-energy $\Sigma_j(i\omega)=\Sigma^{\rm
(0)}_j(\{n_j\})+\Delta\Sigma_j(i\omega)$ {\it but this time} taking
the $n_j$ defining $\Sigma^{\rm (0)}$ as obtained from the lattice
Green function via Eq.~(\ref{njeqn}).  Inserting this self-energy
back into the lattice Green function Eq.~(\ref{latgrneqn}) leads to a
self-consistent condition on both $G_{\bf k}(i\omega)$ and the $n_j$
which may be completely converged with negligible cost, again with
the $\Delta\Sigma_j$ held fixed.  Only after this convergence would
we take the resultant $G_{\bf k}(i\omega)$ and begin the process
of mapping back onto the auxiliary impurity problem in order to
update $\Delta\Sigma_j(i\omega)$.

We found the procedure of converging $\Sigma^{\rm (0)}_j$ in between
each QMC iteration for $\Delta\Sigma(i\omega)$ to work best when the
current QMC ratio $n_{5/2}$:$n_{7/2}$ was retained during the cheap
iterations which were then used only to fix the total $f$ charge
$n_f$.  Also note this approach requires the strict equivalence of
the $n_j$ calculation in both the lattice and impurity problems,
which dictates using
\begin{equation}
G^{\rm imp}_{jm,j^\prime m^\prime}(i\omega) \equiv 
\frac{\delta_{j j^\prime}\delta_{m m^\prime}}{(2j\!+\!1)N}
\sum_{\bf k} Tr_{fj} \{ G_{\bf k}(i\omega) \} \, ,
\label{bath3eqn}
\end{equation}
in conjunction with Eq.~(\ref{slfeeqn}) in solving
Eq.~(\ref{impgrneqn}) for the bath Green function ${\cal
G}_j\delta_{j j^\prime}\delta_{m m^\prime}$.

The output of each QMC cycle is the imaginary time Green function
$\tilde{G}^{\rm imp}_j(\tau)$ which contains statistical
uncertainties, and must be Fourier transformed in order to
define the new self-energy $\Sigma_j(i\omega)$. Following earlier
work,\cite{Held01,McMahan03} our approach is to use multipole fits
\begin{equation}
\tilde{G}^{\rm imp}_j(\tau) \sim \tilde{F}_j(\tau) \equiv
\sum_i w_{ji} \tilde{f}_i(\tau) \, ,
\label{gtfiteqn}
\end{equation}
using basis functions $\tilde{f}_i(\tau)=
-e^{-\varepsilon_i\tau}/(e^{-\beta \varepsilon_i}\!+\!1)$, which have
Fourier transforms $f_i(i\omega)=1/(i\omega\!-\!\varepsilon_i)$,
thus trivially giving the overall Fourier transforms. These fits
were carried out requiring positive weights $w_{ji}\geq 0$, and
constrained to give the QMC values at
$\tau\!=\!0^+$
\begin{equation}
\tilde{F}_j(0^+)  =  \tilde{G}^{\rm imp}_j(0^+) \, 
\label{csn1eqn}
\end{equation}
as well as a number of correct moments $m=1,2,\cdots$
\begin{equation}
\tilde{F}^{(m\!-\!1)}_j(0^+)+\tilde{F}^{(m\!-\!1)}_j(\beta^-) = 
(-1)^m G_j^{(m)} \, .
\label{csn2eqn}
\end{equation}
Here $\beta^- = \beta\!-\!0^+$, $\tilde{F}^{(m)}(\tau) \equiv
(d^m/d\tau^m) \tilde{F}(\tau)$, and the high-frequency moments
$G_j^{(m)}$ are defined by $G^{\rm imp}_j(i\omega) = \sum_m
G_j^{(m)}/(i\omega)^m$.  In our previous work on Ce we imposed
two ($m\!=\!1,2$) of the constraints in Eq.~(\ref{csn2eqn}),
however, we find in the present effort for Pr and Nd at the largest
volumes that it was necessary to add a third moment ($m\!=\!3$).
The reason is that for such strongly localized functions where
the $\tilde{G}^{\rm imp}_j(\tau)$ fall rapidly away from their
$\tau\!=\!0^+$ and $\beta^-$ values, one may get large and unphysical
values of $G_j^{(3)}$ and $G_j^{(4)}$ of opposite signs.  This can
lead to unphysical structure in the spectra far from the chemical
potential, although we could detect no impact on the total energy
or $n_j$ from this problem.  We found that adding the $m\!=\!3$
or $G_j^{(3)}$ constraint was sufficient to fix this difficulty.

The moments $G_j^{(m)}$ may be obtained from {\bf k}-averages and
appropriate traces of powers of $H_{{\rm LDA}}^{0}({\bf k})$ with in
addition $\Sigma_j^{(0)}$ from Eq.~(\ref{sig0eqn}) for $G_j^{(2)}$,
and both $\Sigma_j^{(0)}$ and $\Sigma_j^{(1)}$ for $G_j^{(3)}$.
$\Sigma_j^{(1)}$ may be obtained from the double occupation
matrix,\cite{Oudovenko04} or approximated using a Hubbard-I type
of self-energy tuned to give the QMC $n_j$ values.

The results reported in this paper are for a temperature of 632 K
(0.004 Ry), which given the $T^{-3}$ cost of the QMC calculations
is about as low as is practicable.  Here the imaginary time
interval $(0,\beta)$ for the QMC was discretized by $L\!=\!80$,
$112$, and $160$ divisions, carrying out $10,000$, $6,000$, and
$2,000$ QMC sweeps, respectively, for each DMFT(QMC) iteration.
The smaller number of sweeps for larger $L$ was dictated purely
by the $L^3$ expense of the calculations, and the statistical
uncertainties necessarily increased.  Nonetheless systematic and
generally $L^{-2}$ behavior was seen allowing extrapolation to
eliminate the Trotter errors.\cite{Fye86} Starting self-energies
were already quite good, taken from converged DMFT calculations
with the Hubbard I self-energy,\cite{McMahan03} or from nearby
volumes, or temperatures (0.01 Ry).  At least thirty iterations were
performed at each volume, following five discarded warmup iterations.
At a number of volumes this process was repeated taking the final
self-energy as the initial guess and starting the process from
the beginning.  The results were unchanged to well within the
statistical uncertainties.  All Matsubara sums for an argument
$F(i\omega_n)e^{i\omega_n0^+}$ were carried out using an asymptotic
two-pole approximation $F_{2pol}(i\omega)=w_1/(i\omega-\varepsilon_1)
+w_2/(i\omega-\varepsilon_2)$ with the parameters fixed by
the first four high-frequency moments of $F(i\omega)$.  The
infinite sum was then given by a finite sum over the difference
$F(i\omega_n)-F_{2pol}(i\omega_n)$ plus the analytic result for
the infinite sum over $F_{2pol}$, with $256$ positive Matsubara
frequencies used in the former.

We found a noticeable increase in the QMC statistical uncertainties
for the total energy in going from Ce to Pr to Nd, likely reflecting
the larger role played by the potential energy $U_f d$ and the
QMC determined value of the double occupation, Eq.~(\ref{dbleqn}),
with the increasing number of $4f$ electrons.  At large volumes,
for example, $d$ should be approximately $0$, $1$, and $3$ for Ce,
Pr, and Nd, respectively, at low temperatures.  Similarly we found
differences in the nature of the Trotter corrections, where again at
larger volumes we found the coefficient of the $L^{-2}$ dependence
to be quite small for Ce, larger for Pr, and significantly larger
for Nd.

We were able to get decent spectra from the multipole fits
Eq.~(\ref{gtfiteqn}) by increasing the number of poles by well
over two orders of magnitude. As described earlier,\cite{McMahan03}
we took equally spaced $\varepsilon_i$ grids of $L/4$ points, and
systematically eliminated poles with negative weights, examining
O($10^4$) such grids of varying centroid and width to find the
best fit.  For the spectra we combined the $30$ best fits so long
as the worst of these had a root-mean-square agreement with the QMC
data no more than 20\% larger than that of the best.  More important,
we also averaged over the fits for the last half of the DMFT(QMC)
iterations.  The resultant collections of O($10^3$) poles were
broadened by Gaussians of $0.5$ eV full-width at half maximum.
We found systematic evolution of these spectra with volume providing
one measure of their validity.

\subsection{Double occupation and moment}
\label{dblmomsec}

Since we approximate the $f$-$f$ self-energy by the form
$\Sigma_{jm,j^\prime m^\prime}(i\omega) = \Sigma_j(i\omega)\,
\delta_{j j^\prime} \delta_{m m^\prime}$, a comparable treatment
of the double occupation matrix is
\begin{equation}
\langle \hat{n}_{jm}\hat{n}_{j^\prime m^\prime}
\rangle = \left\{ \begin{array}{ll} 
n_j/(2j\!+\!1) & \mbox{if $j\!=\!j^\prime$, $m\!=\!m^\prime$} \\
d_{jj}/[j(2j\!+\!1)]
& \mbox{if $j\!=\!j^\prime$, $m\!\neq\!m^\prime$} \\
d_{5/2,7/2}/48 & \mbox{if $j\!\neq\!j^\prime$} \\
\end{array} \right. \, , 
\label{n1n2eqn}
\end{equation}
noting that $\hat{n}_{jm}^2\!=\!\hat{n}_{jm}$, and $j\!=\!5/2$
and $7/2$. We obtain $n_j$ and $d_{jj^\prime}$ from the QMC
auxiliary impurity problem by summing over the appropriate blocks
in the $14\times 14$ matrix $\langle \hat{n}_{jm}\hat{n}_{j^\prime
m^\prime}\rangle$.  As these are block totals, the total number
of $f$ electrons per site is $n_f=n_{5/2}\!+\!n_{7/2}$, and the
double occupation $d$ appearing in the expression for the total
energy is $d=d_{5/2,5/2}+d_{5/2,7/2}+d_{7/2,7/2}$.  We find the
present $d_{jj}$ for $j\!=\!j^\prime$ to be bounded above by
their uncorrelated values from $\langle \hat{n}_{jm}\hat{n}_{j
m^\prime}\rangle \sim \langle \hat{n}_{jm}\rangle \langle\hat{n}_{j
m^\prime}\rangle$ for $m\neq m^\prime$, or
\begin{equation}
d_{j}^{\rm unc} = 
j n_j^2 /(2j\!+\!1) \, .
\label{dmaxeqn}
\end{equation}
Similarly, we find the $d_{jj}$ to be bounded below by a typical
strongly correlated expression
\begin{equation}
d_{jj}^{\rm cor} = 
l_j[n_j-(l_j\!+\!1)/2] \, ,
\label{dmineqn}
\end{equation}
where $l_j$ is an integer such that $l_j\!\leq\! n_j\!\leq
\! l_j\!+\!1$ and Eq.~(\ref{dmineqn}) is a piecewise linear function
of $n_j$ with values $n_j(n_j\!-\!1)/2$ at integer $n_j$.

The expectation of the onsite squared $f$ moment is given by
$\langle \hat{J}^2\rangle = 3 \langle \hat{J_z^2}\rangle$ since
our Hamiltonian is rotationally invariant, and thence using
Eq.~(\ref{n1n2eqn}) by
\begin{equation}
\langle \hat{J}^2\rangle = \sum_{j=5/2,7/2}
(j\!+\!1)(jn_j-d_{jj}) \, .
\label{jsqeqn}
\end{equation}
In the local-moment regime at large volume and low temperature
where $n_{5/2}=1,2$, and $3$ for Ce, Pr, and Nd, respectively,
$n_{7/2}\!=\!0$, and $d_{jj}\!=\!n_j(n_j\!-\!1)/2$,
Eq.~(\ref{jsqeqn}) yields $J= 5/2$, $3.2749$, and $7/2$,
respectively, via $\langle \hat{J}^2\rangle=J(J\!+\!1)$.  This
compares to the Hund's rules ground state values of $J_{\rm true}=
5/2$, $4$, and $9/2$, respectively, and reflects the fact that
including the spin-orbit interaction in the absence of intraatomic
exchange will give the correct moment only when the $j\!=\!5/2$
subshell has a single electron or hole, or is trivially empty
or full.  From the thermodynamic perspective, it is the degeneracy
of the Hund's rules multiplet that matters, and this changes from
$14!/[n!(14\!-\!n)!]= 14, 91$, and $364$ to $6!/[n!(6\!-\!n)!]= 6,
15$, and $20$ for Ce, Pr, and Nd, respectively, on adding spin orbit,
which is quite an improvement given the correct degeneracy $2J_{\rm
true}\!+\!1 = 6,9$, and $10$, respectively. Although not described
here, we do find DMFT calculations with the Hubbard-I self-energy
to yield the expected low-temperature entropy plateaus of $\ln 6$,
$\ln 15$, and $\ln 20$ in units of $k_{\rm B}$ for Ce, Pr, and Nd,
respectively, at large volume, similar to earlier work for Ce without
the spin-orbit interaction.\cite{Held01,McMahan03}

One may use Eqs.~(\ref{dmaxeqn}) and (\ref{dmineqn}) to find lower
and upper bounds on $\langle \hat{J}^2\rangle$ in Eq.~(\ref{jsqeqn}).
If one assumes that the spin orbit is also quenched, $n_j=(2j\!+\!1)
n_f/14$, in addition to the uncorrelated Eq.~(\ref{dmaxeqn}), then
\begin{equation}
\langle \hat{J}^2\rangle_{\rm unc} = 
51 \, n_f (1\!-\!n_f/14)/4 \, .
\label{jsqunceqn}
\end{equation}
With this additional assumption, Eq.~(\ref{jsqunceqn}) no longer
provides a lower bound on $\langle \hat{J}^2\rangle$ except at
the smallest volumes where the spin orbit is quenched, but will
nevertheless prove useful.

\begin{figure}[tb]
\centering
\includegraphics[width=3.2in]{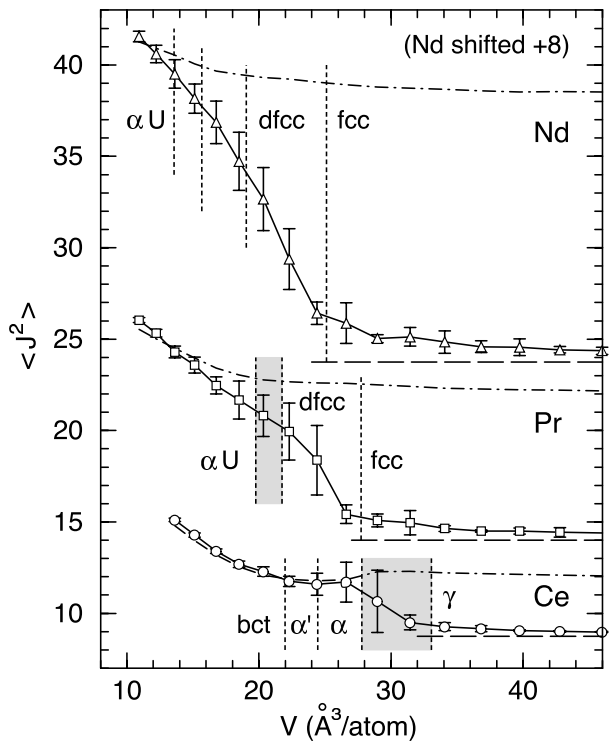}
\caption{Squared local $4f$ moment $\langle J^2 \rangle$ for Ce, Pr,
and Nd. The DMFT(QMC) results are given by the solid curves with data
points; local-moment values, by the horizontal long-dashed lines;
and an uncorrelated estimate with quenched spin orbit, by dash-dot
lines.  The vertical short-dashed lines denote transitions with
significant volume collapse by shading. Both $\alpha$ and $\gamma$
Ce phases are fcc.
\label{jsqfig}}
\end{figure}

\section{Results}
\label{resultsec}

This section reports calculations for compressed Ce, Pr, and
Nd obtained by the LDA+DMFT method with a QMC determination of
the self-energy, to be referred to more simply as just DMFT.
All theoretical results have been carried out for an assumed
fcc structure and at a temperature of 632 K (0.004 Ry) unless
otherwise noted.  This temperature is about as low as is practical
given the $T^{-3}$ expense of the QMC.  Nevertheless, it is cold
enough, since previous work for Ce found that the total energy and
entropy at $632$ K are relatively close to the low-temperature
limit.\cite{Held01,McMahan03} Thus, e.g., the slope of the
$\gamma$--$\alpha$ phase line in Ce comes primarily from the explicit
$T$ in the $TS$ term of the free energy $F=E\!-\!TS$, and not from
$T$-dependence in either the energy or entropy.\cite{Johansson95}

\subsection{Moment and spectra}

The extended aspect of the evolution from localized to itinerant
character in the three lanthanides is illustrated by the DMFT results
(data points with error bars) in Fig.~\ref{jsqfig} for the square
of the onsite $4f$ electron moment $\langle J^2 \rangle$ calculated
using Eq.~(\ref{jsqeqn}).  The vertical dashed lines mark volumes
at which transitions are observed experimentally.  The horizontal
long-dashed lines at large volume give the strongly-correlated
local-moment values $J(J\!+\!1)$ with $J$ the proper Hund's
rules $5/2$ value for Ce, although somewhat smaller than the
proper values for Pr and Nd due to our omission of intraatomic
exchange (see Sec.~\ref{dblmomsec}).  It is apparent for each
material that the local-moment regime persists under compression
up to and through most of the stability field of the fcc phase
(large volume $\gamma$ fcc phase for Ce).  The dash-dot curves
are uncorrelated approximations to $\langle J^2 \rangle$ from
Eq.~(\ref{jsqunceqn}) which also presume that the spin orbit is
quenched, i.e., that $n_{5/2}/n_{7/2} \sim 6/8$, which is largely
responsible for the significant offset of these curves from the
local-moment lines at large volume.  The DMFT results approach
these uncorrelated and spin-orbit quenched curves under compression,
reaching reasonably close agreement by the $\alpha$ phase of Ce and
$\alpha$-U phases of Pr and Nd.  As will be discussed subsequently,
the significance of the quenched spin-orbit lies in the fact that
the lower Hubbard band is of predominant $j\!=\!5/2$ character,
while the $4f$ Kondo resonance which grows at the Fermi level at
the expense of the Hubbard side bands is of mixed $j\!=\!5/2$, $7/2$
character. Therefore quenched spin orbit reflects dominance of the
Kondo peak which may be viewed as a signature of itineracy.

A quantitative measure of the degree of correlation is provided
by the lower (strongly correlated) and upper (uncorrelated)
bounds on the double occupation given by Eqs.~(\ref{dmaxeqn})
and (\ref{dmineqn}), respectively, which in turn provide bounds
on $\langle J^2 \rangle$.  The resultant curves have roughly the
same shape as the DMFT results in Fig.~\ref{jsqfig}, and bracket
these results in each case, with the DMFT curves switching from
more or less perfect agreement with the strongly-correlated
limit at large volume to much closer to the uncorrelated
limit at small volume.  Even so, we find in the vicinity of
$V\sim 12$\AA$^3$/atom that for each of these materials the DMFT
results have switched only about 60\% of the way from the strongly
correlated to the totally uncorrelated limit, consistent with the
fact that we still find residual Hubbard side bands at this volume.
On the other hand, standard paramagnetic LDA certainly does well
enough in predicting the $c/a$ ratio in bct Ce and structural
characteristics of bct and $\alpha$-U phases among the early
actinides,\cite{Brooks84,Soderlind98} so it may well be that for
all practical purposes such phases are weakly correlated enough.

\begin{figure}[bt]
\includegraphics[width=3.2in]{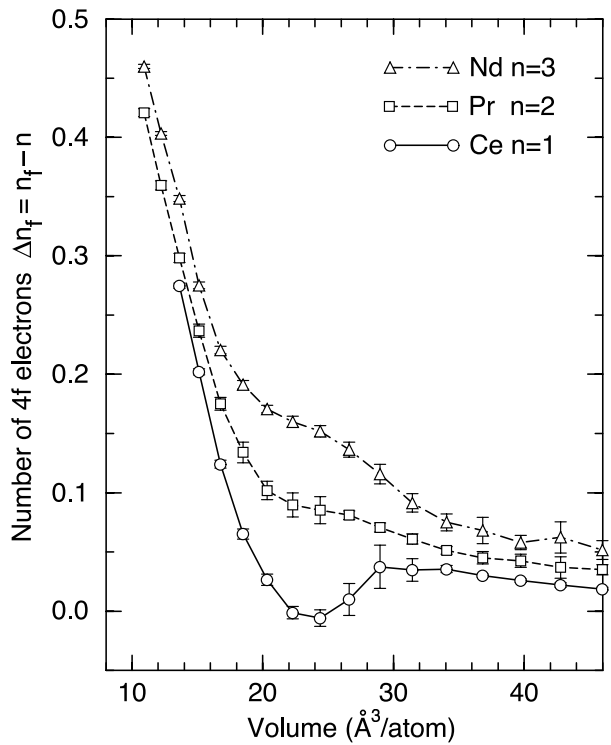}
\caption{Number of $4f$ electrons $n_f$ less an integer $n=1$,
$2$, and $3$ for Ce, Pr, and Nd, respectively.
\label{nffig}}
\end{figure}

\begin{figure}[tb]
\includegraphics[width=3.2in]{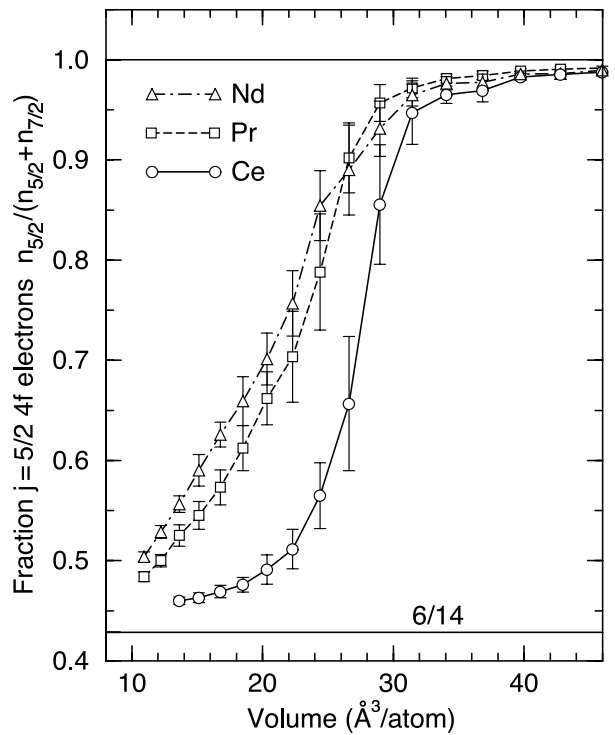}
\caption{ Fraction of $4f$ electrons of $j\!=\!5/2$ character as
a function of volume.
\label{rn1fig}}
\end{figure}

The moment $\langle J^2 \rangle$ in Fig.~\ref{jsqfig} is a bare
quantity and does not reflect any screening effects by the other
electrons.  Its large values at the smaller volumes are due via
Eq.~(\ref{jsqeqn}) to an increase in the number of $4f$ electrons
$n_f$ under compression as seen in Fig.~\ref{nffig}, where $n_f-n$
is plotted with $n=1$, $2$, and $3$ for Ce, Pr, and Nd, respectively.
It is well known that the lanthanides undergo electronic $s$-$d$
transition under pressure, during which the $6s$ and $6p$ states
begin to pass above the Fermi level thereby increasing the $5d$
and $4f$ occupations.\cite{JCAMD} Figure \ref{rn1fig} shows the
fraction $n_{5/2}/n_f$ of $4f$ electrons which are of $j\!=\!5/2$
character, and it is quite clear that there is quenching of the
spin orbit in the region of interest in this work.

\begin{figure}[tb]
\includegraphics[width=3.2in]{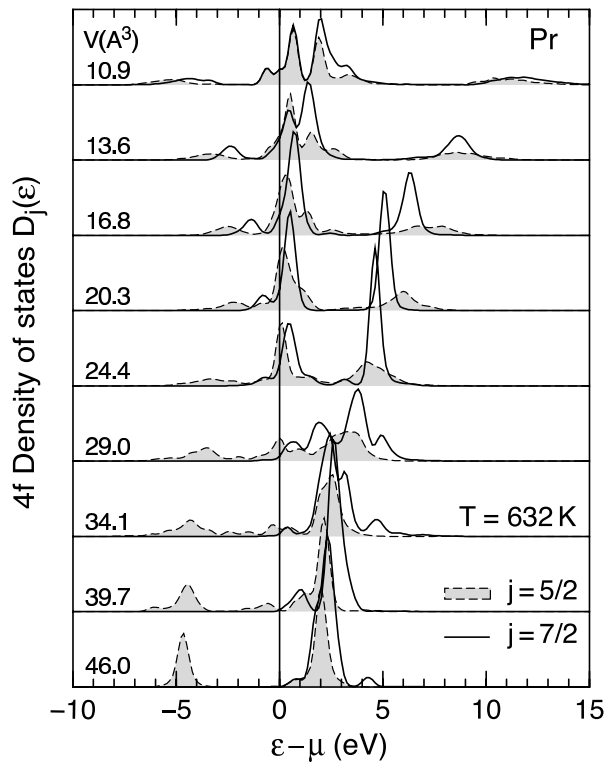}
\caption{Pr $4f$ density of states $D_j(\varepsilon)$ for $j\!=5/2$
(dashed line, shaded) and $j\!=\!7/2$ (solid line) as obtained
from multi-pole fits to the QMC $\tilde{G}_j(\tau)$ with $0.5$
eV FWHM Gaussian broadening.  Areas are normalized $6$:$8$ for
the appropriate $j$'s, and energy is relative to the chemical
potential $\mu$.
\label{prdosfig}}
\end{figure}

The interplay between spin orbit and compression induced changes
in $4f$ spectra may be seen in Fig.~\ref{prdosfig}, where the total
(Brillouin zone summed) but $j$-resolved $4f$ spectra or density of
states $D_j(\varepsilon)$ is plotted for Pr at a number of volumes.
These DMFT results were obtained by multi-pole fits to the QMC
$\tilde{G}_j(\tau)$ as described in Sec.~\ref{cmpdtlsec}, and were
broadened by a Gaussian of $0.5$ eV full width at half maximum.
At the largest volume $V=46.0$ \AA$^3$/atom, one sees the pure
$j\!=\!5/2$ lower Hubbard band near $-5$ eV, and the mixed-$j$
upper Hubbard band near $2$ eV, with the splitting consistent
with $U_f=6.4$ eV at this volume.\cite{prUf} Under compression
both Hubbard bands lose spectral weight at the expense of the
growing Kondo peak near the chemical potential $\mu$, which becomes
dominant at the smallest volumes.  Since this Fermi-level structure
is also of mixed-$j$ character, the population of these states at
the expense of the initially pure $j\!=\!5/2$ lower Hubbard band
relates the quenching of spin orbit in these materials to their
growing itinerant character under compression.  Note, however, as is
especially evident at $V=24.4$ \AA$^3$/atom, that the $j\!=\!5/2$
and $7/2$ contributions to the Kondo peak are split by about the
spin-orbit energy,\cite{Patthey85} with the former more centered
at $\mu$.  Consequently, the occupied part of the Kondo peak is
initially mostly $j\!=\!5/2$ character, although the $j\!=\!7/2$ part
catches up as is also evident from Fig.~\ref{rn1fig}.  The Ce and
Nd spectra are quite similar with the primary visual difference the
increasingly more prominent lower Hubbard band in the progression
from Ce to Pr and Nd, as is to be expected. The Ce spectra are
available elsewhere.\cite{Held01,McMahan03}

A cross check on the spectra in Fig.~\ref{prdosfig} is provided by
\begin{equation} D_j(\mu) \sim (\beta/\pi)\tilde{G}_j(\beta/2)
\, , \label{dosefeqn} \end{equation} where this expression
becomes exact in the low-temperature limit.\cite{Trivedi99}
Equation~(\ref{dosefeqn}) tracks the values at $\varepsilon\!=\!\mu$
in Fig.~\ref{prdosfig} to within better than 20\% at all volumes.
The volume dependence of $D_{5/2}(\mu)$ from Eq.~(\ref{dosefeqn})
looks very much like that plotted for Ce in the case without
spin orbit in Fig.~3 of Ref.~\onlinecite{Held01}.  For decreasing
volume, these values are first quite small then rise to a maximum
signifying growth of the Kondo peak, and then begin to decrease
as the hybridization induced broadening of the $4f$ bands begins
to dominate.  Ce, Pr, and Nd all behave similarly with the onset
of growth in the Kondo peak starting near the low-volume side
of the fcc ($\gamma$-Ce) phases and then rising to successively
higher maxima occurring at smaller volumes, roughly at 29, 23,
and 20 \AA$^3$/atom, respectively.  Comparing these volumes with
the transitions indicated in Fig.~\ref{jsqfig}, it may be seen that
the Ce value reaffirms earlier observations that the initial rapid
rise of the Kondo peak in that material coincides with the volume
collapse.\cite{KVC,KVC2,KVC3,Held01,McMahan03} For Pr and Nd, on the
other hand, the growth of the Kondo peak coincides with the observed
stability field of the dfcc phase, suggesting a transitional role
for this phase rather than being the end member of the localized
trivalent lanthanide series as has been assumed.  In contrast to
the $j\!=\!5/2$ function, $D_{7/2}(\mu)$ starts to increase for
decreasing volume at about the same place as $D_{5/2}(\mu)$, but
does so more gradually, steadily growing until it becomes comparable
to $D_{5/2}(\mu)$ at the smaller volumes.

\subsection{Correlation energy}

\begin{figure}[tb]
\includegraphics[width=3.2in]{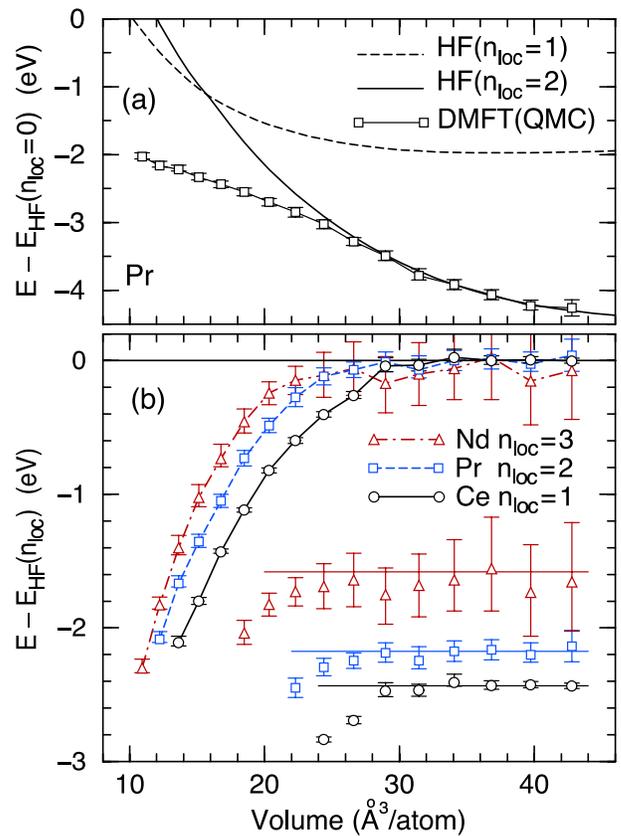}
\caption{(Color online) (a) Correlation energy for Pr as a function
of volume. (b) The more interesting part of the Ce, Pr, and Nd
correlation energies, namely the DMFT energies in each case less
fully polarized Hartree Fock (HF).  The insets (offset vertical
axes) show the large volume behavior more clearly.  The $n_{\rm
loc}$ in $E_{\rm HF}(n_{\rm loc})$ refers to the number of $4f$
bands split off below the Fermi level in the HF calculations, with
$n_{\rm loc}\!=\!0$ the customary paramagnetic solution, $n_{\rm
loc}\!>\!0$ indicating spin- and orbitally-polarized solutions, and
$n_{\rm loc}\!=\!1$, $2$, and $3$ for Ce, Pr, and Nd, respectively,
being the fully polarized solutions.
\label{ecorfig}}
\end{figure}

Figure \ref{ecorfig}(a) shows the Pr correlation energy, namely the
total energy less the result $E_{\rm HF}(n_{\rm loc}\!=\!0)$ for
a paramagnetic Hartree Fock solution of the effective Hamiltonian
Eq.~(\ref{hameqn}).  One may obtain a variety of metastable Hartree
Fock (HF) solutions in which different numbers $n_{\rm loc}$ of
``localized'' $4f$ bands are split off below the Fermi level, ranging
from the paramagnetic solution $n_{\rm loc}\!=\!0$ to the fully
spin- and orbitally-polarized solution (maximum  $n_{\rm loc}$),
which in the case of $4f^2$ Pr is $n_{\rm loc}\!=\!2$.  The $\gamma$
phase of $4f^1$ Ce has been described by analogous fully-polarized
LDA+U solutions ($n_{\rm loc}\!=\!1$),\cite{Sandalov95,Shick01}
and the $\gamma$--$\alpha$ volume collapse by the $n_{\rm
loc}\!=\!1\rightarrow 0$ transition.\cite{Sandalov95}
Fig.~\ref{ecorfig}(a) suggests a two step process in these HF
solutions for Pr, as first one split-off band pops back up to the
Fermi level under compression (the $n_{\rm loc}\!=\!2$ and $1$
curves cross), and then later the second does the same (the $n_{\rm
loc}\!=\!1$ and $0$ curves cross), a $n_{\rm loc}\!=\!2\rightarrow 1
\rightarrow 0$ scenario for which there is no experimental evidence.

On the other hand there is some truth here, as the intermediate
$n_{\rm loc}\!=\!1$ solution is really just a crude attempt to add
$4f$ spectral weight at the Fermi level in the continued presence
of some Hubbard splitting.  For a truly correlated calculation,
in contrast, there is an entirely continuous transfer of spectral
weight from the Hubbard sidebands to the Kondo resonance at the Fermi
level, consistent with the smooth behavior seen in the DMFT curve
in Fig.~\ref{ecorfig}(a).  It agrees with the fully polarized HF
result at large volumes (where the HF correctly captures the Hubbard
splitting), but then bends smoothly away from this HF solution under
compression (where the HF fails to describe the Kondo resonance).
The same behavior has been seen in earlier DMFT results for
Ce,\cite{Held01,McMahan03} as well as in more rigorous QMC solutions
for the Anderson lattice Hamiltonian.\cite{Huscroft99,Paiva03}
It seems intuitively clear that such deviation from the fully
polarized HF solution as volume is reduced should be associated
with the growth of $4f$ spectral weight at the Fermi level, i.e.,
the Kondo resonance.

The total energy differences between DMFT and fully polarized HF
solutions of Eq.~(\ref{hameqn}) are shown in Fig.~\ref{ecorfig}(b)
for fcc Ce, Pr, and Nd.  Ferromagnetic order was assumed for
the HF.  The three insets (shifted vertical axes) give a clearer
view of the large volume behavior, and the systematically larger
QMC uncertainties from Ce to Pr to Nd have been discussed in
Sec.~\ref{cmpdtlsec}.  Our calculations of the $4f$ spectra indicate
growth of the Kondo resonance for each material first begins on the
small-volume side of the experimentally observed fcc or $\gamma$-Ce
stability range (see Fig.\ref{jsqfig}), and then $D_{5/2}(\mu)$
reaches a maximum at $\sim 29$, $23$, and $20\,$\AA$^3$/atom for
Ce, Pr, and Nd, respectively.  These volumes are close to where the
curves in Fig.~\ref{ecorfig}(b) begin to bend away from the fully
polarized HF (0 baseline), consistent with the association of this
energy with the Kondo resonance.

The fact that these deviations occur at successively smaller
volumes from Ce to Pr to Nd is consistent with the ratio of the
Coulomb interaction to band width $U_f/W_f$ for the $4f$ states,
which near equilibrium volume ($\sim 34\,$\AA$^3$/atom) is 3.7, 4.4,
and 4.9 for Ce, Pr, and Nd, respectively.\cite{JCAMD} Thus Pr and
Nd are successively more localized than Ce at ambient conditions,
and greater compression is needed to bring about similar changes in
electron correlation.  A region of negative bulk modulus is of course
essential to obtain a first-order isostructural transition as in Ce.
If more generally a region of low bulk modulus were to favor larger
volume changes in structural phase transitions, then the delayed
response of the softening effects in Fig.~\ref{ecorfig}(b) for
Pr and Nd means competing against the ever more dominant positive
bulk moduli of the remaining contributions to the total energy as
they continue to grow with decreasing volume.  This is at least
intuitively consistent with a 15\% volume collapse in Ce, 9\% in Pr,
and apparently none in Nd.  It is difficult to quantify this point
as the QMC uncertainties in Fig.~\ref{ecorfig}(b) don't admit very
reliable calculation of the curvature of these results.

There are also some important differences between the Ce and Pr
volume collapse transitions, beyond the fact that the former
is isostructural and the latter involves a structural change.
As already noted, the experimentally observed two-phase region
(27.8--$33.1\,$\AA$^3$/atom) for the $\gamma$--$\alpha$ collapse
in Ce coincides with the region of rapid grown in the Kondo
resonance.  It also overlaps a rather sharp breakaway of the
DMFT curve from the HF result, as seen near $29\,$\AA$^3$/atom
in Fig.~\ref{ecorfig}(b), which is consistent with a large negative
contribution to the bulk modulus.  Here, as for earlier results on Ce
without the spin orbit,\cite{Held01,McMahan03} as well as analyses
of the Anderson lattice Hamiltonian,\cite{Huscroft99,Paiva03}
these features corroborate the Kondo-volume collapse model of
the Ce $\gamma$--$\alpha$ transition.\cite{KVC,KVC2,KVC3} For
Pr, on the other hand, the volume collapse appears to occur at
pressures above the region of rapid growth of the Kondo resonance.
The maximum in the Pr $j\!\!=5/2$ spectral weight at the Fermi
level is near $23\,$\AA$^3$/atom (see also Fig.~\ref{prdosfig}),
and the Pr curve in Fig.~\ref{ecorfig}(b) begins to bend away from
the fully polarized HF at slightly larger volume, both of which
lie outside and to the large-volume side of the observed two-phase
region for Pr (19.8--$21.8\,$\AA$^3$/atom).  If the Ce collapse
serves to initiate the evolution from localized to itinerant in
this material, such evolution is already well underway by the Pr
collapse, a point which is also quite evident from behavior of the
local moment in Fig.~\ref{jsqfig}.

\subsection{Equation of state}

\begin{figure}[tb]
\includegraphics[width=3.2in]{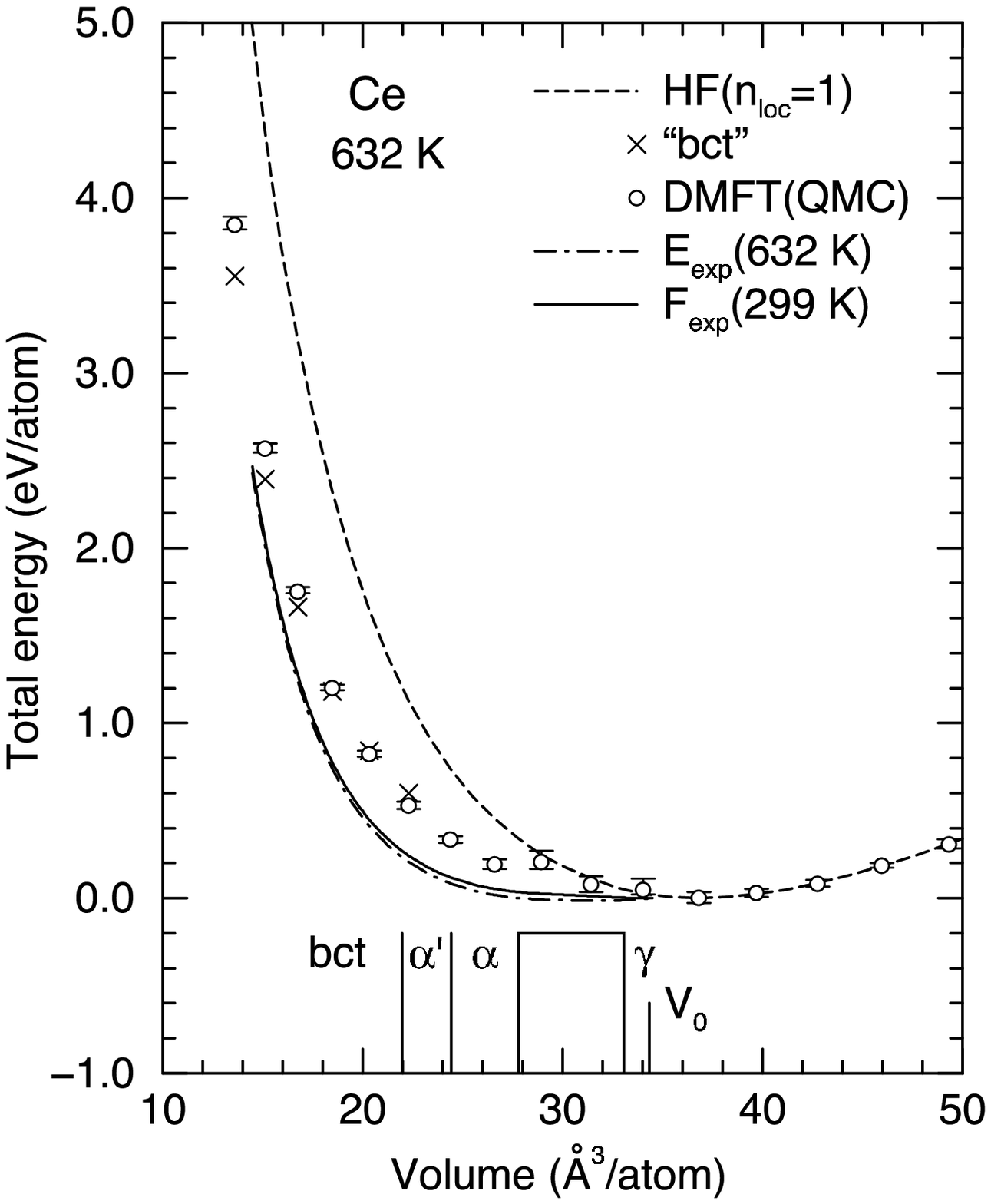}
\caption{Total energy of Ce as a function of volume.
The theoretical HF and DMFT results are at 632 K.  The
experimental energy (632 K) and free energy (299 K) are obtained
from Refs.~\onlinecite{VohraCe99} and \onlinecite{SchiwekCe02} as
described in the text.  The theoretical bct estimate took a $T\!=\!0$
local-density bct--fcc energy difference,\cite{RavindranCe98}
and added this to the present fcc DMFT result. The experimentally
observed 300-K equilibrium volume $V_0$ and stability fields for
various phases are marked.
\label{etotcefig}}
\end{figure}

\begin{figure}[tb]
\includegraphics[width=3.2in]{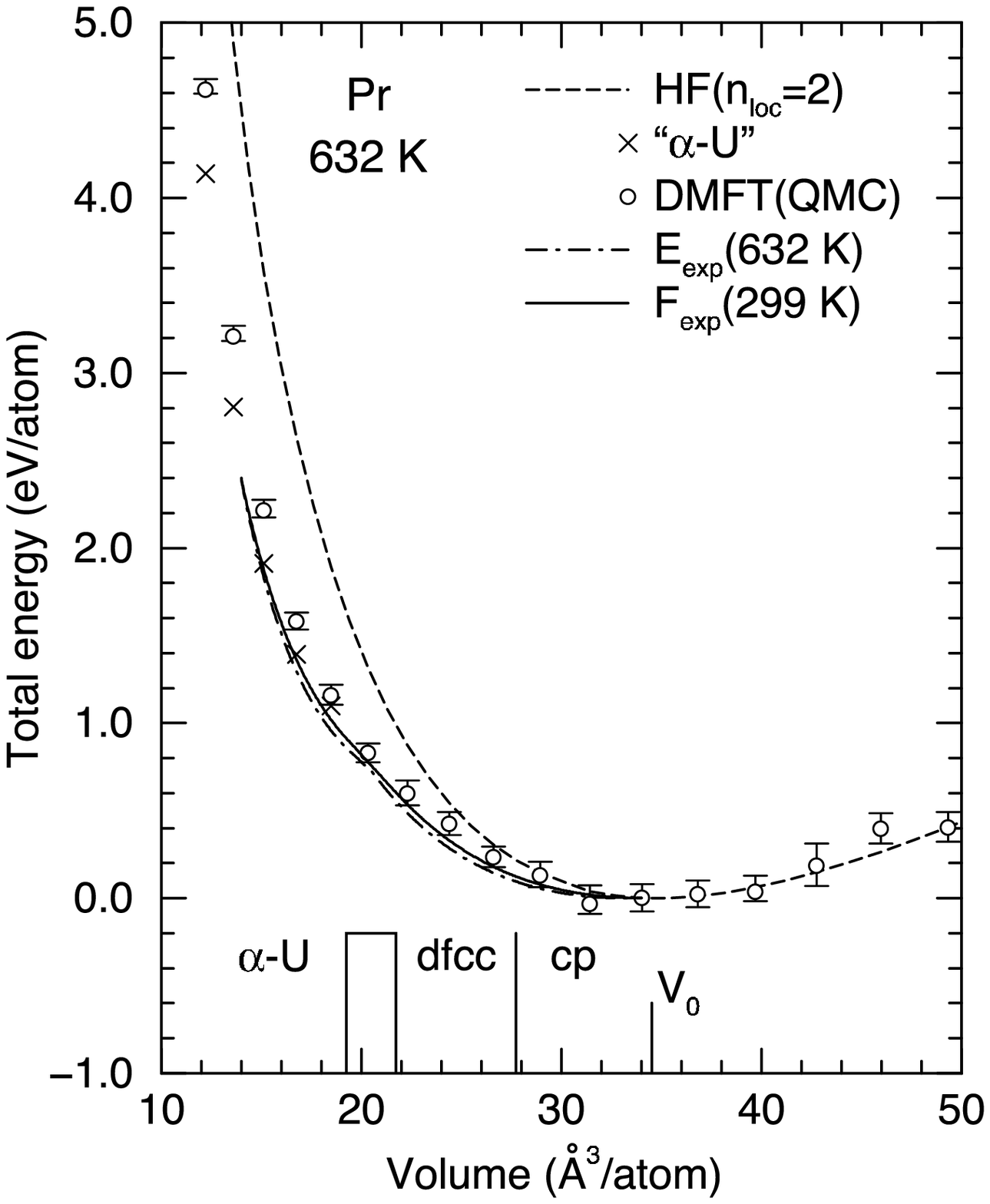}
\caption{Total energy of Pr as a function of volume.
The theoretical HF and DMFT results are at 632 K.
The experimental energy (632 K) and free energy (299 K) are
obtained from Ref.~\onlinecite{BaerPr03} as described in the text.
The theoretical $\alpha$-U estimate took a $T\!=\!0$ local-density
$\alpha$-U--fcc energy difference,\cite{SoderlindPr02} and added
this to the present fcc DMFT result. The experimentally observed
300-K equilibrium volume $V_0$ and stability fields for various
phases are marked (cp denotes close packed).
\label{etotprfig}}
\end{figure}

\begin{figure}[tb]
\includegraphics[width=3.2in]{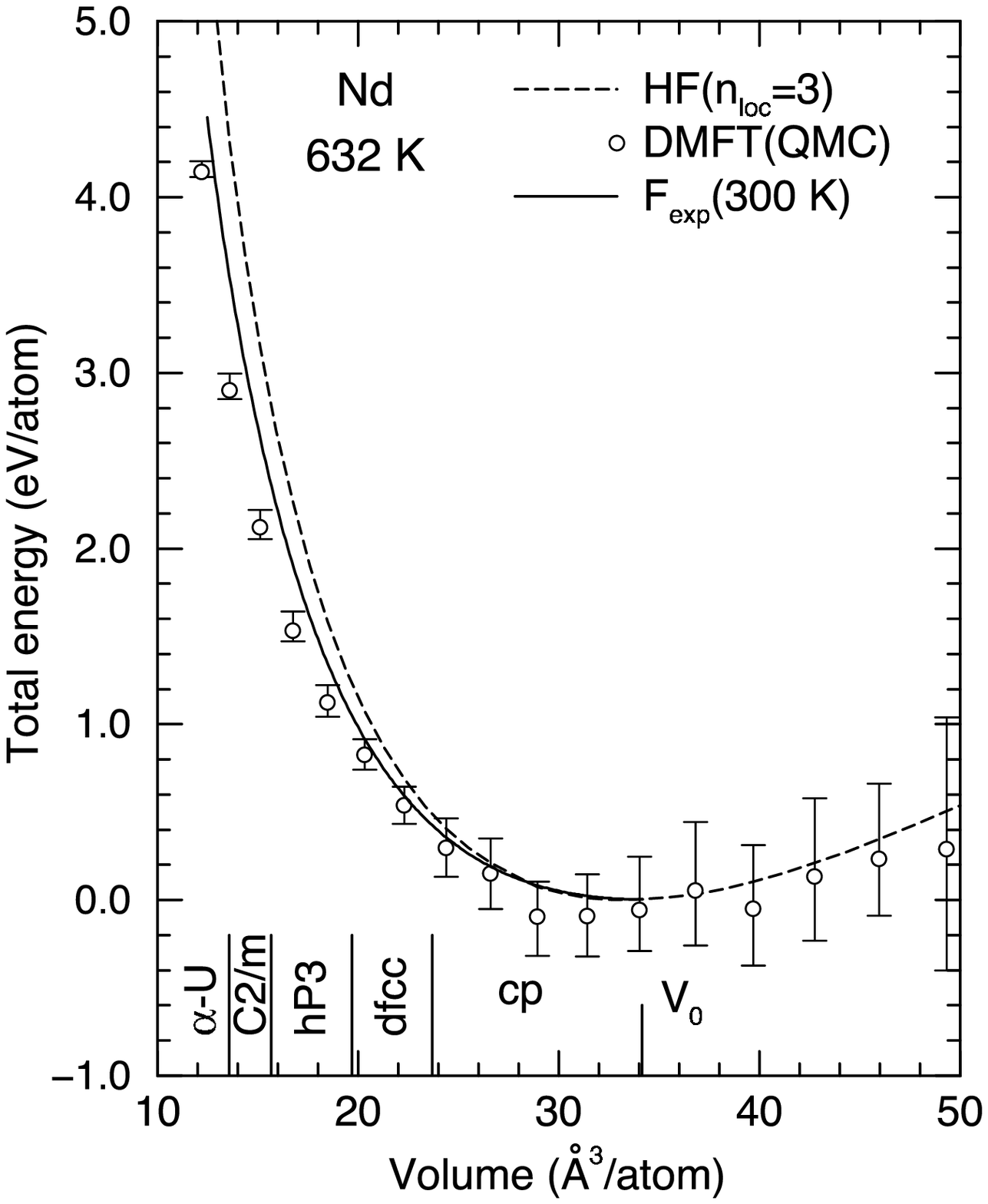}
\caption{Total energy of Nd as a function of volume.  The theoretical
HF and DMFT results are at 632 K.  The experimental free
energy (300 K) was obtained from Ref.~\onlinecite{ChesnutNd00}. The
experimentally observed 300-K equilibrium volume $V_0$ and stability
fields for various phases are marked (cp denotes close packed).
\label{etotndfig}}
\end{figure}

Comparison of the theoretical and experimental energies are given
for Ce, Pr, and Nd in Figs.~\ref{etotcefig}, \ref{etotprfig}, and
\ref{etotndfig}, respectively.  The HF and DMFT total energies
are all at 632 K and for an assumed fcc structure as noted.
Experimental pressure data $P(V,T)$ exists for Pr over a wide range
of volume at $300 < T < 725$ K,\cite{BaerPr03} from which one may
obtain the free energy $F(V,T)$ and total energy $E(V,T)$ to within
arbitrary constants, which we fix by setting $E(V_0,T)=F(V_0,T)=0$
at the experimental 300-K equilibrium volume $V_0$.  Raising the
temperature generally makes the free energy a steeper function of
volume.  However, since $\partial P/\partial T= \partial S/\partial V
> 0$, the total energy is generally less steep than the free energy.
It is a fortunate coincidence that these effects roughly cancel so
that the 300-K free energy is within 0.07 eV/atom of the 632-K total
energy for Pr throughout the range plotted for these experimental
quantities in Fig.~\ref{etotprfig}, noting again that both are zeroed
at $V_0$.  Experimental measurements of $P(V,T)$ for Ce exist up 208
GPa at 300 K,\cite{VohraCe99} and over the temperature range 299--573
K within the fcc ($\gamma$ and $\alpha$) regime.\cite{SchiwekCe02}
Assuming $\partial P/\partial T$ is independent of volume, we may
use the former results to extend the latter up to higher pressures.
Extrapolating these results also to somewhat higher temperatures,
one may obtain the $F(V,299\,$K$)$ and $E(V,632\,$K$)$ curves seen in
Fig.~\ref{etotcefig} which like Pr are also fairly close.  We presume
that it is also legitimate to compare room-temperature free energies
for Nd to our 632-K calculated total energies for that material.

The Debye temperatures for Ce, Pr, and Nd are all below 150 K
at ambient conditions,\cite{Gschneidner64} and with reasonable
Gr\"uneisen parameters all three materials should still be in
the high-temperature limit at 632 K throughout the volume range
studied here.  The phonon contribution to the total energy is then
$3k_{\rm B}T$ and has no impact on the volume dependence examined
in Figs.~\ref{etotcefig}--\ref{etotndfig}.

The total energies in Figs.~\ref{etotcefig}--\ref{etotndfig} provide
another perspective on the correlation issues discussed earlier.
The fully-polarized HF results (dashed) curves are analogous to
LDA+U and should do well at large volumes in the strongly localized
limit as is the case here.  The DMFT results (open circles) then
show how such static mean field theories begin to break down as
volume is reduced due to their inability to account for the growing
Kondo resonance and its contribution to the correlation energy.
This effect is most pronounced for Ce in Fig.~\ref{etotcefig},
less so for Pr in Fig.~\ref{etotprfig}, and smallest for Nd in
Fig.~\ref{etotndfig}, reflecting the fact that Pr and Nd start out
successively further in the localized limit than Ce, therefore
requiring greater compression to achieve comparable changes in
the correlation energy.  Although the agreement between the DMFT
results and the experimental isotherms is not ideal, and will be
discussed further, it is clear that experiment does confirm this
systematic progression.

A comparison of theoretical and experimental values for the bulk
properties is given in Table~\ref{bulktab}.  The theory results
were obtained from 6-term fits \cite{Birch52} to the energies
over the range $17$--$49\,$\AA$^3$/atom in order to better average
over scatter in the QMC results.  Table~\ref{bulktab} shows quite
decent agreement with experiment,\cite{Delin98} and is a partial
validation of the present effective Hamiltonians Eq.~(\ref{hameqn})
and the manner of total energy calculation Eq.~(\ref{etoteqn}).
Of particular note is the fact that the DMFT results for $V_0$
and $B_0$ are in better agreement with experiment and 3\% and
36\% smaller, respectively, than the fully polarized HF values
in the case of Ce, as compared to $\sim\!1$\% and $6$--$9$\%
smaller, respectively, for Pr and Nd. This suggests a small but not
unimportant effect of the Kondo resonance even at $P\!=\!0$ in Ce,
effects which are shifted to smaller volume and have less impact
for Pr and Nd.

\begin{table}
\caption{ Comparison of theoretical and experimental values
for the equilibrium volumes $V_0$(\AA$^3$/atom) and bulk
moduli $B_0$(GPa).  The theory used fits to the 632-K results
in Figs.~\ref{etotcefig}--\ref{etotndfig}, while the experiment
is at 300 K.  However, the Ce HF results change by only $0.4$\%
and $-2.6$\% for $V_0$ and $B_0$, respectively, on reducing the
temperature from 632 to 316 K.}
\begin{ruledtabular}
\begin{tabular}{ccccccc}
& $V_0^{\rm HF}$ & $V_0^{\rm DMFT}$ & $V_0^{\rm exp}$\footnotemark[1]
& $B_0^{\rm HF}$ & $B_0^{\rm DMFT}$ & $B_0^{\rm exp}$\footnotemark[2] \\
\hline
Ce & $37.0$ & $35.8$ & $34.37$ & $33.3$ & $21.2$ & $20\,$--$\,21$ \\
Pr & $34.9$ & $34.5$ & $34.54$ & $34.2$ & $31.0$ & $26\,$--$\,37$ \\
Nd & $33.5$ & $33.0$ & $34.18$ & $34.9$ & $32.9$ & $28\,$--$\,32$ \\
\end{tabular}
\end{ruledtabular}
\footnotetext[1]{From Ref.~\onlinecite{Gschneidner90}.}
\footnotetext[2]{From Ref.~\onlinecite{Grosshans92}.}
\label{bulktab}
\end{table}

An estimate of the structural contribution to the energy at smaller
volumes is provided by the ``bct'' and ``$\alpha$-U'' results
($\times$ symbols) in Figs.~\ref{etotcefig} and \ref{etotprfig} for
Ce and Pr, respectively.  These were obtained by adding paramagnetic
local-density functional values for the bct--fcc energy difference
in Ce,\cite{RavindranCe98} and the $\alpha$-U--fcc difference in
Pr,\cite{SoderlindPr02} to the DMFT fcc values.  Throughout the
ranges shown in Figs.~\ref{etotcefig} and \ref{etotprfig} these
structural energy differences are 10\% or less for Ce and 25\% or
less for Pr of the more dominant correlation energy contribution
represented by the separation between the DMFT and the polarized
HF curves.  These structural corrections improve the agreement with
experiment somewhat for Ce and Pr, however, a similar correction
for Nd would move the DMFT results to lower energies further away
from the experiment.

\begin{figure}[tb]
\includegraphics[width=3.2in]{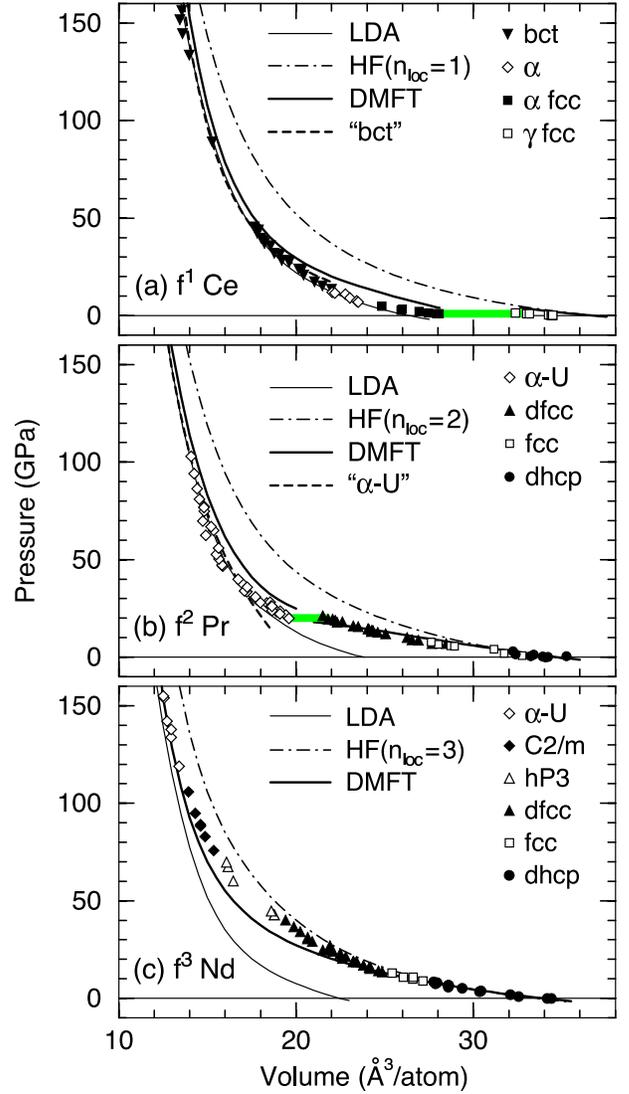}
\caption{(Online color) Experimental 300-K pressure-volume
data for (a) Ce,\cite{KoskimakiCe78,OlsenCe85,VohraCe99}
(b) Pr,\cite{MaoPr81,ZhaoPr95,ChesnutPr00,BaerPr03} and (c)
Nd,\cite{ChesnutNd00,AkellaNd86,AkellaNd99} as compared to theory.
The observed volume collapse transitions in Ce and Pr are shaded.
The thin solid lines give $T\!=\!0$ LDA results,\cite{presASA} while
the HF, DMFT, and structural estimates (``bct'' Ce, ``$\alpha$-U''
Pr) are negative volume derivatives of the total energies in
Figs.~\ref{etotcefig}--\ref{etotndfig}.
\label{prsfig}}
\end{figure}

Figure \ref{prsfig} gives the pressure-volume results corresponding
to Figs.~\ref{etotcefig}--\ref{etotndfig}, with experimental 300-K
data shown for Ce,\cite{KoskimakiCe78,OlsenCe85,VohraCe99}
Pr,\cite{MaoPr81,ZhaoPr95,ChesnutPr00,BaerPr03} and
Nd.\cite{ChesnutNd00,AkellaNd86,AkellaNd99} Except for the
$T\!=\!0$ LDA results (thin solid lines),\cite{presASA} the
theoretical results are negative volume derivatives of the 632-K
total energies in these figures, an approximation justified by the
$F_{\rm exp}(300\,{\rm K})\sim E_{\rm exp}(632\,{\rm K})$ comparisons
in Figs.~\ref{etotcefig} and \ref{etotprfig}.  Multi-term fits to
the energies were used in obtaining the derivatives.\cite{Birch52}
The DMFT points (open circles) were fit ignoring the error bars,
with separate fits made for the regions above and below the observed
collapse transitions for Ce and Pr.  It is evident that the present
paramagnetic DMFT results do well in both low-pressure localized
and high-pressure itinerant extremes, especially considering the
structural corrections [dashed lines in Figs.~\ref{prsfig}(a) and
(b)] in the later regime, although there are evident discrepancies
in between as will be discussed.  Standard paramagnetic LDA does
well in the itinerant regime as seen by the thin black lines, but
must be combined with one of the magnetically-ordered modified-LDA
techniques (e.g., LDA+U) to capture the larger volume behavior.
This then leads to a magnetic order-disorder transition which is
not observed at room temperature, and, e.g., an incorrect prediction
of a volume collapse (16\% at 39 GPa) in Nd.\cite{Eriksson90}

Turning to the collapse transitions, the kink in the DMFT total
energy for Ce near $29\,$\AA$^3$/atom in Fig.~\ref{etotcefig}
lies within the experimental two-phase region, is within the
QMC error bars, and is similar to earlier results without spin
orbit.\cite{Held01,McMahan03} The addition of the spin-orbit
interaction has lowered the $\gamma$-phase energy more than that
of the $\alpha$ phase, so that a common tangent construction to the
DMFT energy curve in Fig.~\ref{etotcefig} gives a pressure of $4\pm2$
GPa as compared to a slightly negative value in the earlier work.
While the experimental transition ranges from (an extrapolated)
$-0.7$ GPa at $T\!=\!0$ to 1.8 GPa at 485 K,\cite{SchiwekCe02} an
error of 4 GPa is not huge on the scale of Fig.~\ref{prsfig}(a).
The volume change at 4 GPa in the DMFT results (thick solid lines)
in Fig.~\ref{prsfig}(a) is in decent agreement with that observed
at room temperature.  In the case of Pr, the estimated $\alpha$-U
curve (dashed line) in Fig.~\ref{prsfig}(b) is in good agreement
with experiment over all but the lowest observed range of this
phase. This raises the question of whether the structural energy
difference may be critical to the Pr collapse.  Indeed, while we
find a softening in the Pr equation of state for the fcc structure
due to the contribution in Fig.~\ref{ecorfig}(b), there is no
convincing evidence that this drives the bulk modulus negative in
the vicinity of the collapse.

Finally, we turn to the discrepancies between theory and
experiment in Figs.~\ref{etotcefig}--\ref{prsfig}.  The fact
that the DMFT results for Ce in Fig.~\ref{etotcefig} are too
high at smaller volumes in comparison with experiment, those
for Pr in Fig.~\ref{etotprfig} are reasonable, while those for
Nd in Fig.~\ref{etotndfig} are somewhat low is suggestive of the
intraatomic exchange interaction, $K$, which has been omitted in
the present work.  In the strongly localized regime including $V\sim
V_0$, this exchange interaction should shift the energy by $0$, $-K$,
and $-3K$ for Ce, Pr, and Nd, respectively.  At small volumes, the
itinerant expectation would be $-K\sum_\sigma \sum_{m\!<\!m^\prime}
\langle n_{m\sigma} n_{m^\prime \sigma}\rangle \sim -3Kn_f^2/14$
taking $\langle n_{m\sigma} n_{m^\prime \sigma}\rangle\sim
(n_f/14)^2$.  If the total energies are shifted to $0$ at $V_0$
as done in these figures, this would suggest shifts of $-0.31$,
$0.08$, and $0.69$ eV at $V\!=\!15\,$\AA$^3$/atom for Ce, Pr,
and Nd, respectively, taking values of $n_f$ from Fig.~\ref{nffig}
and $K\sim 1\,$eV.  This would imply completely itinerant states
at $15\,$\AA$^3$/atom which is unlikely especially for Nd.  Indeed,
polarized HF calculations suggest exchange corrections at this volume
of $-0.13$, $0.07$, and $0.25\,$eV,\cite{hfexch} respectively,
although this also is only an estimate.  Nevertheless, these
estimates do suggest that intraatomic exchange would significantly
improve the present comparisons between theory and experiment.
Unfortunately, a rigorous inclusion of the exchange interaction,
one that would also yield the the correct Hund's rules values
for the Pr and Nd local moments, would require DMFT calculations
for the full $f$-$f$ Coulomb interaction including all four
Slater integrals and the non-density-density exchange and pair
hopping terms.  This has been a challenge for QMC, although there
is recent progress.\cite{Sakai04} Similarly, implementing mutual
self-consistency between the LDA and DMFT parts of the calculations,
as has been advocated,\cite{Savrasov04} would be prohibitively
expense with the present QMC implementation of the self-energy.
Both improvements as well as DMFT calculations of the structural
energy differences could be considered with faster although less
rigorous approaches to the self energy.\cite{Savrasov05}

\section{Summary and Discussion}
\label{summarysec}

A better understanding of the electron-correlation driven volume
collapse transitions in the compressed lanthanides may come from
putting this behavior in context of their extended evolution from
localized to itinerant character.  To this end, the present paper has
reported calculations for compressed Ce ($4f^1$), Pr ($4f^2$), and
Nd ($4f^3$) using a combination of the local density approximation
(LDA) and dynamical mean field theory (DMFT), so called LDA+DMFT.
Results for the $4f$ moment $\langle \hat{J^2}\rangle$, spectra,
correlation energy, and equation of state among other quantities
have been presented over a wide range of volume at a temperature
of 632 K. This temperature is the lowest feasible with the
present quantum Monte Carlo (QMC) implementation of the self
energy, yet is still reasonably close to the low-temperature
limit.\cite{Held01,McMahan03} While a face-centered cubic (fcc)
structure was assumed, LDA estimates of the important structural
energy differences are significantly smaller than the relevant
contributions to the correlation energy which may then be described
as leading order volume-dependent effects.

We find the three lanthanides to remain rather strongly localized
under compression from ambient conditions up through the observed
stability fields of the fcc ($\gamma$ Ce) phases, in the sense that
the $4f$ moments are close to the Hund's rules values, there are
fully formed Hubbard sidebands which are themselves a signature of
the local moments, and little $4f$ spectral weight lying in between
at the Fermi level.  Subsequent compression brings about significant
deviation of the moments from the Hund's rules values, growth of
$4f$ spectral weight at the Fermi level (the Kondo resonance) at
the expense of the Hubbard sidebands, an associated softening in the
total energy, and quenching of the spin orbit given that the Kondo
peak is of mixed-$j$ character in contrast to the predominantly
$j\!=\!5/2$ lower Hubbard band.  The most dramatic evolution in
these signatures is seen to coincide with the two-phase region of
the $\gamma$--$\alpha$ phase transition in the case of Ce, consistent
with earlier work,\cite{Zoelfl01,Held01,McMahan03} and in agreement
with the Kondo volume-collapse scenario.\cite{KVC,KVC2,KVC3} For
Pr and Nd, on the other hand, these signatures change most rapidly
over the volume range where the distorted fcc (dfcc) structure is
experimentally observed to be stable, suggesting that this phase
is transitional and not part of the localized trivalent lanthanide
sequence.

Only on subsequent compression is Pr experimentally observed to
undergo a collapse from the dfcc phase to an $\alpha$-U structure,
while Nd passes from dfcc to $\alpha$-U through two other low
symmetry phases without any substantial volume changes.  Due to the
increasing but incompletely screened nuclear charge, the lanthanides
shift towards the localized limit for larger atomic number, and
we see a similar off-set to smaller volume of the above mentioned
signatures from Ce to Pr, and then Nd.  In particular, a softening
contribution to the total energy associated with growth of the Kondo
resonance must compete with the remaining contributions which become
ever more dominant due to a steadily increasing bulk modulus as
volume is reduced.  If a region of low bulk modulus were to favor
larger volume changes in structural transitions, then this would be
qualitatively consistent with the observed sequence 15\% (Ce), 9\%
(Pr), and none (Nd).  This speculation is apparently contradicted by
the 5\% and 6\% collapse transitions in Gd ($4f^7$) and Dy ($4f^9$),
respectively.  However, these lanthanides correspond to filling a
different spin-orbit subshell, which may significantly complicate
matters given the profound manner in which spin-orbit is involved.

There are still some correlation effects evident at even the
smallest volumes considered here, such as residual Hubbard side
bands in the $4f$ spectra and moments whose values have evolved only
$\sim 60$\% of the way from the strongly localized Hund's rules
values to those characteristic of totally uncorrelated electrons.
Nevertheless, standard paramagnetic LDA does quite well within
the bct phase of Ce and the $\alpha$-U phases of Pr and Nd, and so
for all practical purposes these phases appear weakly correlated
enough. The success of LDA calculations for the structural parameters
in these phases for Ce as well as the light actinides is also well
known.\cite{Brooks84,Soderlind98}

This work has also included a detailed quantitative comparison
between the present LDA+DMFT results and experiment for the
total energies and pressures of Ce, Pr, and Nd.  The comparison is
encouraging, and serves to corroborate such theoretical observations
as the systematic offset of Kondo-like correlation signatures to smaller
volume for increasing atomic number in these three lanthanides.
There are also clear deficiencies, most notably the need to include
intraatomic exchange, ideally in its full rotationally invariant
form to enable rigorous calculation of the Hund's rules moments in
general cases.  The need for accurate LDA+DMFT structural energy
differences is also apparent.  These improvements as well as making
the LDA and DMFT parts mutually self-consistent will require a
self-energy treatment that is both more precise and considerably
less expensive, which unfortunately is likely to require giving up
some of the rigor of the present QMC approach.

\section*{Acknowledgments}
This work was performed under the auspices of the U.S. Department
of Energy by the University of California, Lawrence Livermore
National Laboratory under contract No. W-7405-Eng-48.  The author
gratefully acknowledges extended interactions with Stewardship
Science Academic Alliances collaborators at U.C. Davis (DOE/NNSA
Grant NA00071), especially R.T Scalettar.  He has benefited from
conversations with K.~Held, and with members of the DOE/BES funded
Computational Materials Science Network Cooperative Research Team
on ``Predictive Capability for Strongly Correlated Systems.''
A modification of the QMC code of Ref.~\onlinecite{DMFT2} (App.~D)
was used in the present work.


\begin{thebibliography}{99}
\bibitem{softenergy}  
A rapid drop in entropy across the $\gamma$--$\alpha$
collapse in Ce may be viewed via $S(V,T)=\int_0^T dT^\prime
C_V(V,T^\prime)/T^\prime$ as arising from a specific heat $C_V(V,T)$
peak which lies below, e.g., $T\!=\!300\,$K on the larger-volume
$\gamma$ side, but moves above $300\,$K on the smaller-volume
$\alpha$ side.  The finite-$T$ total energy, $E(V,T)=E_0(V)+\int_0^T
dT^\prime C_V(V,T^\prime)$, should then show a concomitant softening
near the collapse, possibly even a region of negative curvature,
even though it is the free energy which dictates the transition, and
regardless of whether the $T\!=\!0$ energy $E_0(V)$ is featureless.
\bibitem{cubirr}
In the full cubic point group, $\Sigma(i\omega)$ would be
characterized by five unique diagonal matrix elements corresponding
to the irreducible representations $\Gamma_6^-$ ($j\!=\!7/2$),
$\Gamma_7^-$ ($5/2$ \& $7/2$), $\Gamma_8^-$ ($5/2$ \& $7/2$),
with unique off-diagonal couplings between the two $\Gamma_7^-$
and between the two  $\Gamma_8^-$.
\bibitem{prUf}
The Hubbard $U_f$ values for Ce, Pr, and Nd are plotted in
Ref.~\onlinecite{JCAMD}, and, e.g., for Pr are $6.4$, $5.9$,
and $4.7$ eV for $V=$ $46.0$, $29.0$, and $10.9$ \AA$^3$/atom,
respectively.
\bibitem{presASA}
These are LMTO ASA calculations including the combined and Ewald
corrections for an fcc structure.
\end{thebibliography}
\end{document}